\documentclass[a4paper]{revtex4-2}

\usepackage[english]{babel}
\usepackage[utf8]{inputenc}
\usepackage[labelfont=up]{subcaption}
\usepackage[x11names]{xcolor}
\usepackage{float}
\usepackage{mathtools}
\usepackage[top=1.in, bottom=1.in, left=0.5in, right=0.5in]{geometry}
\usepackage{amsthm, amssymb}
\usepackage{graphicx}
\usepackage{lipsum}
\usepackage{bm}
\usepackage[pdftex, pdftitle={Article}, pdfauthor={Author}]{hyperref}
\usepackage{wasysym}
\usepackage{enumitem}
\hypersetup{colorlinks=true, citecolor=black, linkcolor=black, urlcolor=blue}

\DeclarePairedDelimiter\ton{(}{)}
\DeclarePairedDelimiter\qua{[}{]}
\DeclarePairedDelimiter\gra{\{}{\}}

\DeclarePairedDelimiter\mean{\langle}{\rangle}

\begin{document}
\title{Recommender systems may enhance the discovery of novelties}
\author{Giordano De Marzo$^{1, 2, 4, 5}$, Pietro Gravino$^{2, 3, 4}$ and Vittorio Loreto$^{1, 3, 4, 5}$}
\affiliation{
$^1$Sapienza University of Rome, Physics Department and Sapienza School for Advanced Studies, P.le A. Moro, 2, 00185, Rome, Italy.\\
$^2$Sony Computer Science Laboratories Paris, 6, Rue Amyot, 75005, Paris, France\\
$^3$Sony Computer Science Laboratories Rome, Joint Initiative CREF-SONY, Piazza del Viminale, 1, 00184, Rome, Italy.\\
$^4$Centro Ricerche Enrico Fermi, Piazza del Viminale, 1, 00184, Rome, Italy.\\
$^5$Complexity Science Hub Vienna, Josefstaedter Strasse 39, 1080, Vienna, Austria.
}

\date{\today} 

\begin{abstract}
Recommender systems are vital for shaping user online experiences. While some believe they may limit new content exploration and promote opinion polarization, a systematic analysis is still lacking. We present a model that explores the influence of recommender systems on novel content discovery. Surprisingly, analytical and numerical findings reveal these techniques can enhance novelty discovery rates. Also, distinct algorithms with similar discovery rates yield varying opinion polarization outcomes. Our approach offers a framework to enhance recommendation techniques beyond accuracy metrics.
\end{abstract}
\maketitle

\section{Introduction}
Innovation plays a crucial role in human life. It represents how our society progresses~\cite {johnson_2010_book,sood2010interacting, rzhetsky2015choosing, youn2015invention}. On the other hand, novelties, meant as innovations at the individual level, also enter our everyday life when we listen to a new song, watch a new movie, or try a new recipe~\cite{lu2012recommender, wagner2021measuring}. Novelties and innovations share an essential feature: they can be viewed as first-time occurrences at the individual or collective level, respectively~\cite{tria2014dynamics}. 

The question remains open about how individuals and communities explore and access new content. If, in the past, we primarily relied on suggestions by friends or experts for becoming aware of new content and experiences, nowadays, recommendation algorithms play a central role in selecting the content we are exposed to and thus also the novelties we encounter. Indeed, we rely on algorithmic recommendations for choosing series to watch, music to listen to on streaming platforms, or items to buy on online marketplaces. Given the ubiquity of personalized recommendations, since the pioneering work of Pariser~\cite{pariser2011filter} introducing the concept of Filter Bubble, much attention has thus been devoted to the study of recommendation algorithms, with the aim of better understanding their effect on individuals and society as a whole. Indeed, if, on the one hand, recommendations allow us to access content otherwise lost in the immensity of the Internet, on the other side, they could potentially constrain us in algorithmic bubbles that strongly limit the diversity of what we see~\cite{nagulendra2014understanding, dillahunt2015detecting, gravino2019towards, Perra2019, DeMarzo2020, kirdemir2021exploring, Peralta2021, peralta2021opinion, santos2021link, iannelli2022filter, cinus2022effect, bellina2023effect, valensise2023drivers}. Despite a sizeable scientific activity in this area, analysis has yet to be performed to understand recommendation algorithms' potential effects on discovering novelties. Though those algorithms suppress, in principle, the exploration of new items due to the so-called popularity bias, they also favour novelties thanks to the collaboration among users they induce. It is thus essential to understand which of these two factors is dominating. It is well known that discovery processes are typically described by Heaps' law, which relates the number of distinct elements $D(t)$ experienced to the total number of elements $t$ by a sublinear scaling $D(t)\sim t^{\beta}$. This law has been observed in numerous systems~\cite{youn2015invention, lu2010zipf, de2021dynamical, marzo2022modeling, di2023dynamics}, including online platforms~\cite{tria2014dynamics}. It is thus natural to question if it is affected by the presence of recommendation algorithms and how they may alter the scaling exponent $\beta$.

To address this issue, in the present letter, we propose a model of content exploration under the effect of recommendation algorithms. Our framework allows us to compare individual exploration and exploration under the different recommendation schemes, understanding to what extent they limit or enhance the ability to experience novelties. Analytical and numerical results suggest that recommendation algorithms increase the rate of novelty discovery, even though they may augment polarization. 

\section{Results}
\subsection{The Model}
Let us consider a set of $N$ users and $M$ possible different items, such as songs, movies, etc. The users are affected by a user-user collaborative filtering algorithm that determines their propensity to choose one item rather than the other. Such a technique, one of the most famous and widespread, is based on the idea that users with similar preferences or behaviours in the past will have similar preferences in the future. On this basis, it recommends items to a user based on the opinions or ratings of other like-minded users. Thus, if we denote by $r_{ui}$ the rating the user $u$ assigns to an item $i$, the user-user collaborative filtering algorithm estimates the probability $R_{ui}$ for a generic user $u$ to like the item $i$ as~\cite{resnick1994grouplens, lu2012recommender}:
\begin{equation}
		R_{ui}=\frac{\sum_v^N S_{uv}r_{vi}}{\sum_v^N S_{uv}}.
		\label{eq:collaborative_filtering_general}
\end{equation}
Here, $S_{uv}$ is the similarity between users $u$ and $v$, estimated starting from the ratings $r_{ui}$. Many different definitions are possible. In this paper, we use the cosine similarity, defined as follows, but other choices give analogous results.
\begin{equation}
S_{uv}=\frac{\sum_i r_{ui}r_{vi}}{\sqrt{\sum_i r_{ui}^2}\sqrt{\sum_i r_{vi}^2}}.
\end{equation}
In most situations, the ratings are not directly available because, generally, users do not provide explicit feedback on the item they interact with~\cite{rendle2021item}. For instance, considering a music platform, the engagement of a user $u$ with a given song $i$ is estimated from the number of times $N_{ui}$ the user has previously listened to the song $i$, as follows:
\begin{equation}
r_{ui}=\frac{\rho N_{ui}+1}{\sum_i^{M_u}(\rho N_{ui}+1)}=\frac{\rho N_{ui}+1}{\rho t_u + M_u},
\end{equation}
where $t_u=\sum_i N_{ui}$ is the total number of songs previously listed by the user, $\rho$ is a reinforcing parameter setting the relevance we assign to plays, and $M_u$ is the number of songs we are considering in the recommendations. The term $+1$ we are adding to $\rho N_{ui}$ ensures that the algorithm might recommend never played songs and avoids the algorithm's "cold start"~\cite{schein2002methods}. In general, $M_u$ is evolving since new songs are composed and new artists enter the scene. Following Kauffman's concept of adjacent possible~\cite{kauffman1996investigations}, we adopt the same mechanism proposed in the Urn Model with Triggering (UMT)~\cite{tria2014dynamics, tria2018zipf}. Every time a novelty occurs, it triggers the expansion of the adjacent possible, i.e., of the space of items that, though not yet selected, are just a step away from being so. This conditional mechanism implies that each time a user listens to a novel song, $M_u$ gets increased by a term $\nu+1$, mimicking the fact that the novel listening action triggers other potential new songs the user may like that were not being previously considered in the recommendations. Denoting by $D_u$ the number of distinct songs played by $u$, it thus holds $M_u=(\nu+1)D_u+M_0$, with $M_0$ being the number of songs available at time zero. We also assume that users follow the same path in exploring the adjacent possible, meaning that if the number of distinct elements $D_u$ clicked by the user $u$ is larger than those clicked by $v$, $D_v$, then the adjacent possible of $v$ is a subset of that available to $u$. Putting everything together, we can thus rewrite Eq.~\eqref{eq:collaborative_filtering_general} as 
\begin{equation}
		R_{ui}=\frac{\sum_v^N S_{uv}\theta\qua*{(\nu+1)D_v+M_0-i}\frac{\rho N_{vi}+1}{\rho t_u + (\nu+1) D_v + M_0}}{\sum_v^N S_{uv}}.
		\label{eq:collaborative_filtering_urn}
\end{equation}
The probability of randomly selecting a specific user $u$ for the update is given by $R_{ui}/N$. In the above expression, $\theta$ denotes the Heaviside step function; it takes into account the fact that if $i$ is larger than $(\nu+1)D_v+M_0$, quantifying the number of distinct songs available to the user $v$, then item $i$ is not yet available to $v$, which thus do not contribute to the probability of recommending such a song. Having defined the recommendation probability, the model runs by selecting at each time step $\delta t=1/N$ a random user and then making it click on an item $i$ with a probability given by Eq.~\eqref{eq:collaborative_filtering_urn}. Note that for $N=1$, i.e., for a single user, the model reduces to the standard UMT~\cite{tria2014dynamics}. Therefore, our model can also be seen as a set of $N$ urns of the UMT type with a common path in the adjacent possible and interacting through a recommendation algorithm. This is unsurprising since the user-user collaborative filtering expression Eq.~\eqref{eq:collaborative_filtering_general} is just a Polya urn when only one user is considered.

\begin{figure}[t]
    \includegraphics[width=\linewidth]{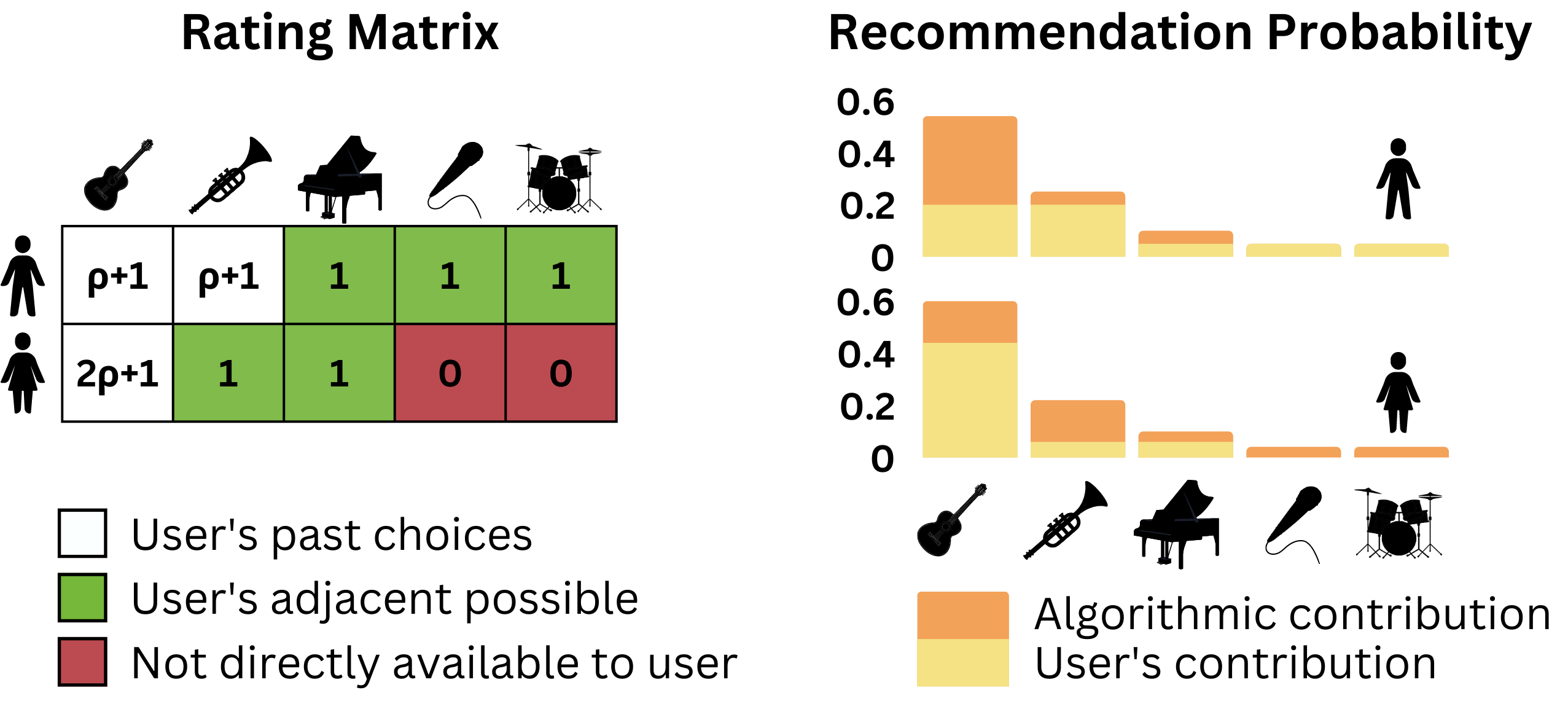}
    \caption{\textbf{Recommendation Probability.} We show the rating matrix on the left for recommending musical instruments to two users with $\nu=1$. White boxes correspond to musical instruments previously listened to by the users, green ones to those never listened to but contained in users' adjacent possibles. In contrast, red boxes denote those genres that would be inaccessible without a recommendation algorithm. The male users clicked once on two distinct instruments, and thus his adjacent possible contains three other musical instruments. At the same time, the female one listened twice to the same instrument, so her adjacent possible is smaller. On the right, we report the recommendation probabilities for the specific value $\rho=3$ that gives a similarity between users of $0.78$. We show each user's contribution from its listening history (yellow) and the other user via the recommendation algorithm (orange).}
    \label{fig:scheme_model}
\end{figure}

We provide an example of how the recommendation algorithm works in Fig.~\ref{fig:scheme_model}, where we show a rating matrix for the simple case of two users and the corresponding recommendation probabilities. As it is possible to see, while in the single-user case, the first two musical instruments would be clicked with the same probability by the male user, collaborative filtering gives a probability boost to the first instrument since it has been listened to twice by the female user. Analogously, even if the latter has no direct access to the last two instruments, they become available thanks to the presence of the male user and the recommendation algorithm.
\begin{figure}[t]
    \includegraphics[width=0.95\linewidth]{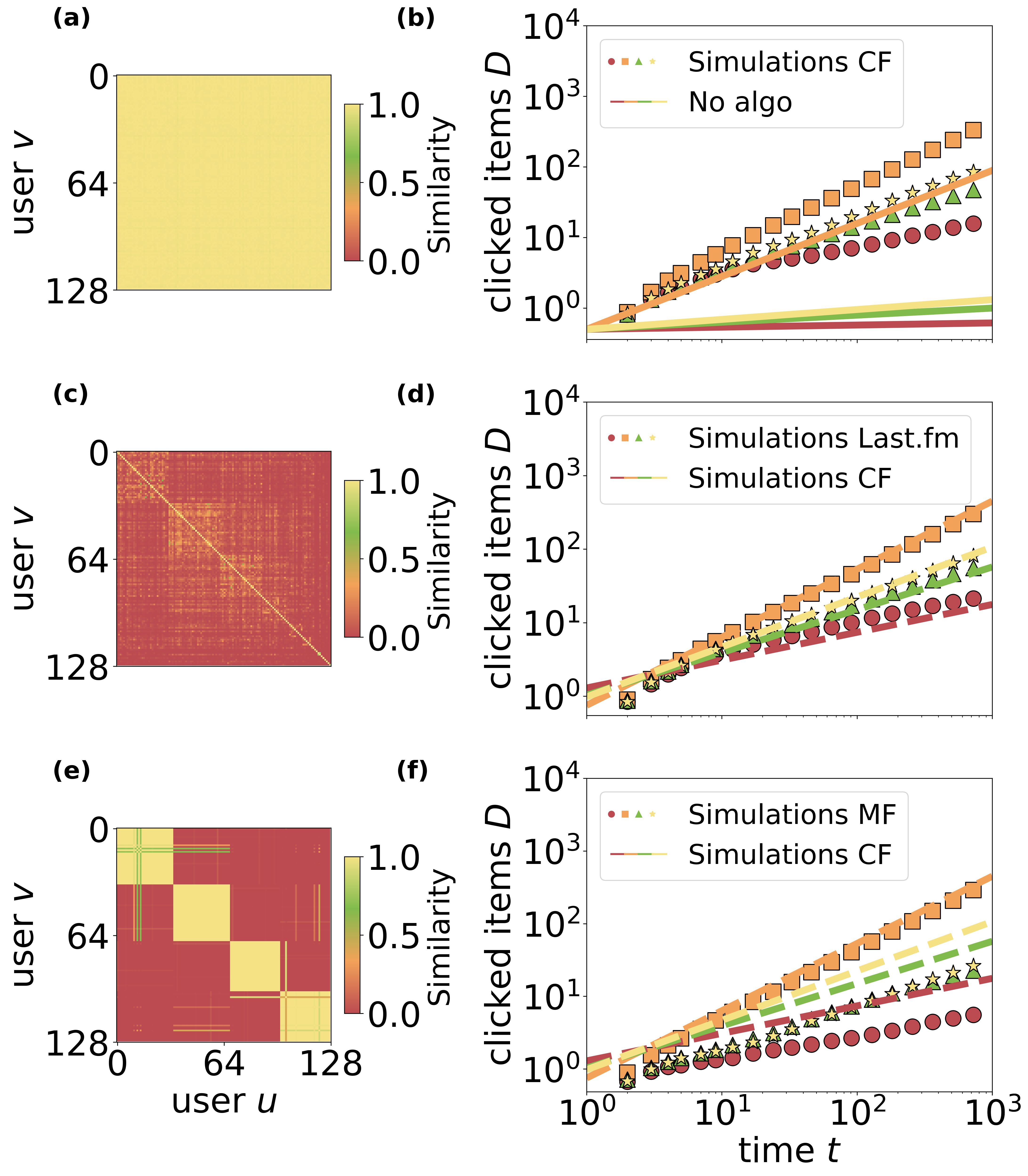}
    \caption{\textbf{Numerical simulations.} \textbf{a)} Asymptotic similarity matrix of our modeling scheme for the following values of the parameters: $M_0=4$, $N=128$, $\nu=3$, and $\rho=100$. The similarity among users tends to the unitary value, leading to a consensus configuration. \textbf{b)} Evolution of the model's average number of distinct items as a function of time (scatter plots). We used $M_0=4$, $N=128$ and $\nu=3, \rho=100$ (red circles), $\nu=3, \rho=4$ (orange squares), $\nu=5, \rho=50$ (green triangles), $\nu=7, \rho=50$ (yellow stars). We also report the behaviour without a recommendation algorithm (solid lines) corresponding to the same values of $\nu$ and $\rho$. \textbf{c)} Similarity matrix of $128$ users from the online music platform Last.fm. Rows and columns have been ordered, exploiting greedy modularity maximization. \textbf{d)}  As in \textbf{b)}, using the real similarity matrix from Last.fm to run the user-user collaborative filtering. Dashed lines show the time evolution of $D$ given by our model (panel b). \textbf{e)} Asymptotic similarity matrix of our model when the matrix factorization is adopted for the following values of the parameters: $M_0=4$, $N=128$, $\nu=3$, and $\rho=100$. Rows and columns have been ordered by exploiting greedy modularity maximization. A modular structure corresponding to a polarized configuration is present. \textbf{f)} As in \textbf{b)}, but using the matrix factorization algorithm to run the model. Dashed lines show the time evolution given by our model. The matrix factorization shows the same Heaps' law, though with a smaller prefactor.}
    \label{fig:numerical_simulations}
\end{figure}

\subsection{Numerical Simulations}
We wish to investigate two aspects of recommender systems: their propensity to enhance polarization and their ability to boost content exploration. The former is controlled by the similarity matrix they induce. A block structure in the similarity matrix would imply that users are split into groups characterized by different tastes and behaviours. We show in Fig.~\ref{fig:numerical_simulations} the asymptotic similarity matrix for a particular choice of the parameters of our modeling scheme (the same results are obtained for other values of the parameters.) In this case, users reach a consensus configuration where $S_{uv}\approx1$, meaning that $r_{ui}\approx r_{vi}$ for all $u, v$. The first conclusion is that user-user collaborative filtering, without other mechanisms, such as social interaction or personal tastes, does not produce polarization. As a second step, to assess the effect of the user-user collaborative filtering on novelties' discovery, we look at the evolution of the average number of distinct elements, $D=\mean*{D_u}$, experienced by users. We report the time evolution of $D$ in Fig.~\ref{fig:numerical_simulations}b, where $D$ exhibits follow Heaps' law, i.e., a power law growth in time of the number of distinct elements experienced. The $\beta$ exponent of the power-law is independent of $M_0$ and $N$, provided that $N$ is large enough. For each combination of the parameters $\nu$, $\rho$, we also report the behaviour as a solid line of $D$  when no collaborative filtering mechanism exists. In this case, one observes a smaller value of Heaps' exponent, i.e., a lower rate of novelty discovery. This result implies that the user-user collaborative filtering increases the pace of novelty discovery.
As we saw above, the model we introduced, guided by the collaborative filtering scheme, always leads to a consensus configuration where all similarities tend to be one. In reality,  users have different tastes influencing their choices and more complex social dynamics, leading to the emergence of distinct communities rather than a complete consensus. Consequently, similarity matrices derived from the behaviour of users on online platforms can be very different from those generated by our model. To test this point, we considered the listening histories of $128$ real users on the online music platform Last.fm, which we used to reconstruct their similarity matrix. Launched in 2002, Last.fm is a music discovery platform that provides personalized music recommendations. It builds a detailed profile of each user's musical tastes by recording details of the tracks the user listens to on various streaming platforms or devices. Through analyzing these data, Last.fm employs collaborative filtering algorithms that identify shared listening patterns among users to generate tailored playlists and recommendations. Last.fm data have already been analyzed in several studies~\cite{konstas2009social, tria2014dynamics, kowald2020unfairness}, and they represent a sort of standard for music recommender systems. In particular, we focus on the Music Listening Histories Dataset (MLHD), which contains more than 27 billion time-stamped logs extracted from Last.fm~\cite{vigliensoni17music}. To build a rating matrix from the dataset, we selected $128$ random users and filtered out artists that received less than $1000$ plays. We then defined the entry $N_{ui}$ as the number of times the $u$-the user listened to the $i$-th artist. This allows us to define the ratings $r_{ui}$ and consequently the similarity matrix $S_{uv}$. This matrix, depicted in Fig.~\ref{fig:numerical_simulations}c, is then exploited in the numerical simulation as the $S_{uv}$ appearing in Eq.~\eqref{eq:collaborative_filtering_urn}, keeping it fixed during the simulation. Results are shown in Fig.~\ref{fig:numerical_simulations}d. Again we observe a faster pace of novelties thanks to the presence of a recommendation algorithm and no substantial differences to the previous cases (reported as dashed lines). This result shows that the precise structure of the similarity matrix does not play a significant role in determining Heaps' exponent, which is mainly affected by the parameters $\nu$ and $\rho$.
 
Finally, we need to investigate the generality of the previous results for different recommendation techniques. To this end, we considered a different, more sophisticated, and more realistic recommendation algorithm, i.e., the matrix factorization~\cite{ma2008sorec, he2016fast}. Such an algorithm aims to predict users' preferences or ratings for items by factorizing the user-item rating matrix $r$ into two lower-rank matrices, $U$ and $V$. This factorization helps capture latent features or patterns in the data. It is obtained by minimizing the error between the predicted ratings (achieved by multiplying $U$ and $V$) and the observed ratings in the user-item matrix. This approach is one of the most widely used in real implementations of recommender systems because it elaborates accurate recommendations in reasonable delays, even for large systems, thanks to the approximation implied by the dimensionality reduction. The dimensionality of the latent space is, in fact, a key factor in the optimization, even if its role in users' opinion dynamics is largely unknown. In this work, we exploited the function NMF (non-negative matrix factorization) available in the Python machine learning library Scikit learn REF. At each time, we select a random user $u$ and input the user-item rating matrix $r$ in the NMF, which returns the predicted user-item matrix $\tilde{r}$. Such a matrix is then used to recommend the user $u$ an item $i$ with probability $P_i=\tilde{r}_{ui}/\sum_i\tilde{r}_{ui}$. The first important result is that the matrix factorization gives rise to polarized configurations, meaning that users spontaneously split into distinct groups with strong inner and low intergroup similarity. This behaviour is shown in Fig.~\ref{fig:numerical_simulations}e, where we report the similarity matrix for the case $M_0=4$, $N=128$, $\nu=3$ and $\rho=100$. Rows and columns have been ordered to highlight the different communities using greedy modularity maximization. We then show in Fig.~\ref{fig:numerical_simulations}f the evolution of novelties under the NMF, compared with the user-user collaborative filtering simulated with the same parameters (dashed lines). Also, in this case, we observe the same power-law behaviour, but, as an effect of the modular structure, the scaling prefactor of the matrix factorization is lower.  
    
\subsection{Analytical Results}
This section provides an analytical grounding for the results reported above. We first look at how the similarity among users evolves. The master equation governing the evolution of $N_{ui}$. Denoting by $Q(N_{ui}, t)$ the probability of the number of clicks at time $t$, one has:
 \begin{align*}
            Q(N_{ui}, t+\delta t) =& R_{ui}\ton*{N_{ui}-1}Q(N_{ui}-1, t)+\\
            &+[1-R_{ui}\ton*{N_{ui}}]Q(N_{ui}, t)
\end{align*}
For large $N$ (small $\delta t$) we can rewrite the master equation as
\[
    \frac{dQ(N_{ui})}{dt}=\delta tR_{ui}\ton*{N_{ui}-1}Q(N_{ui}-1, t)-\delta tR_{ui}\ton*{N_{ui}}Q(N_{ui}, t).
\]
We can then compute the time derivative of the average value $\mean*{N_{ui}(t)}$ as
\begin{align*}
    \frac{d\mean*{N_{ui}(t)}}{dt}=\sum_{N_{ui}=0}^{\infty}N_{ui}(t)\frac{dQ(N_{ui})}{dt}.
\end{align*}
This gives:
\begin{equation}
    \frac{d\mean*{N_{ui}(t)}}{dt}\approx \sum_{N_{ui}=0}^{\infty}Q(N_{ui})R_{ui}\ton*{N_{ui}}=\mean*{R_{ui}}
    \label{eq:drift}
\end{equation}
        
It is useful to introduce the normalized ratings $n_{ui}$ defined as 
\[
    n_{ui}=\frac{N_{ui}}{t}
\]
which satisfies the asymptotic martingale property and thus has an asymptotic limit. By using Eq.~\eqref{eq:drift} we can write the temporal evolution of their mean values as
\[
    \frac{d\mean*{n_{ui}}}{dt}=\frac{1}{t}\qua*{\mean*{R_{ui}}-\mean*{n_{ui}}}
\]
and if we neglect stochastic fluctuations removing the average value, we end up with a system of coupled differential equations of the form
\begin{equation}
    \frac{dn_{ui}}{dt}=
    \frac{1}{t}\qua*{R_{ui}-n_{ui}}
    \label{eq:evolution_norm_ratings}
\end{equation}
Here we can use the expression for the clicking probability introduced in Eq.~4 of the main text
\begin{equation*}
		R_{ui}=\frac{\sum_v^N S_{uv}\theta\qua*{(\nu+1)D_v+M_0-i}\frac{\rho N_{vi}+1}{\rho t_u + (\nu+1) D_v + M_0}}{\sum_v^N S_{uv}}.
            \label{eq:collaborative_filtering_urn}
\end{equation*}
Focusing on an item $i$ already available to all users, we rewrite the evolution of the normalized ratings as
\[
    \frac{dn_{ui}}{dt}=
    \frac{1}{t}\qua*{\frac{\sum_v^NS_{uv}\ton*{n_{vi}-n_{ui}}}{\sum_v^NS_{uv}}},
\]
where we also took the limit of large $t$. Setting the time derivative to zero, we see that only two classes of stable stationary states are possible:
\begin{itemize}
    \item consensus with $n_{ui}=n_{vi}$ for all items $i$ and all pairs of users $u, v$;
    \item polarized solution with $S_{uv}=0$ for all pairs of users such that $n_{ui}\neq n_{vi}$.
\end{itemize}
However, only the former is an attractor of the dynamics, as can be seen by considering a simple case with $N=2$. For the similarity between the two users to be null, it must be $n_{1i}>0$ if and only if $n_{2i}=0$, so suppose we initialize the system slightly away from this configuration:
\[
    n_{1i}=k_{1i}-\frac{\epsilon}{M_d} \text{for } i\leq M_1 \text{ and }n_{1i}=\frac{\epsilon}{M_d} \text{for }i> M_d
\]
\[
    n_{2i}=\frac{\epsilon}{M_d} \text{for } i\leq M_d \text{ and }n_{2i}=k_{2i}-\frac{\epsilon}{M_d} \text{for }i> M_d.
\]
We can then study how $\epsilon$ evolves in time to understand if the dynamics moves the system toward the polarized solution or not. Without lack of generality, let us consider the first user and $i>M_d$. We have
\[
    \frac{dn_{1i}}{dt}=\frac{1}{M_d}\frac{d\epsilon}{dt}=\frac{1}{t}\qua*{\frac{S_{12}\ton*{n_{2i}-n_{1i}}}{S_{12}}}=\frac{k_{2i}-2\epsilon}{t}.
\]
Since $\epsilon$ is a small quantity compared to $k_{2i}$, its time derivative is positive, and so it increases up to the value $\epsilon=k_{2i}/2$, which corresponds to $n_{1i}=n_{2i}$ and so to consensus. This argument, which can be easily generalized to any $N$, proves that the system spontaneously evolves toward a configuration in which all users equally rate the different items so that $S_{uv}$ goes to one in the large time limit independently of $u$ and $v$.
Summarizing, these results imply that the only possible stationary state of the system corresponds to all users clicking with the same frequency on the different items $n_{ui}\approx n_{vi}$. Consequently, as we saw in the numerical simulations, the similarity asymptotically tends to one $\lim_{t\to\infty}S_{uv}\approx1$.
 
\begin{figure}[t]
    \includegraphics[width=\linewidth]{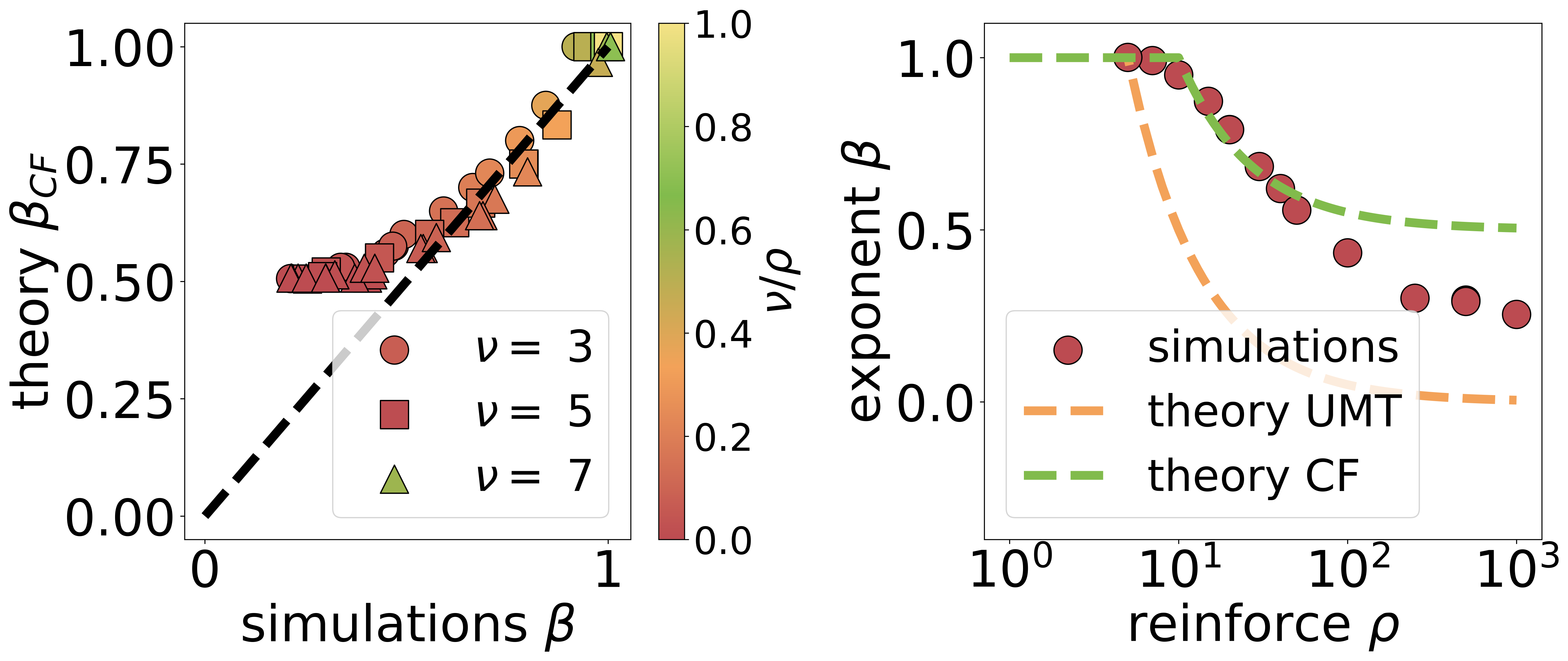}
    \caption{\textbf{Theoretical predictions.} Left panel. Comparison between the Heaps' exponent observed in numerical simulations and the corresponding theoretical prediction for different values of $\nu, \rho$ and $M_0=1$, $N=128$, $T=10^4$. The ratio $\nu/\rho$ is given by the colour bar, with red corresponding to this quantity going to zero. The dashed line is the bisector. Theory and simulations agree when $\nu/\rho$ is large, while discrepancies arise when the ratio decreases. Right panel. Heaps' exponent as function of $\rho$ for $\nu=5$, $M_0=1$, $N=128$, $T=10^4$ in numerical simulations (points). We also show the theoretical prediction (green dashed line) and the behaviour without recommendation algorithms (orange dashed line). Numerical simulations show that, as we predicted, Heaps' exponent does not tend to zero as $\rho$ increases, but the asymptotic value is smaller than expected.} 
 \label{fig:theoretical_predictions}
\end{figure}

We can now move to the study of novelties occurring in the system. The probability for users to click on items is given by Eq.~\eqref{eq:collaborative_filtering_urn} and reads:
\[
    R_{ui}=\frac{1}{N}\frac{\sum_v^N S_{uv}\theta\qua*{(\nu+1)D_v+M_0-i}\frac{\rho N_{vi}+1}{\rho t_u + (\nu+1) D_v + M_0}}{\sum_v^N S_{uv}}
\]
As mentioned above, in the large time limit, the system reaches a consensus configuration, so we can assume $S_{uv}=1$ and write the transition rate as
\[
    R_{ui}=\frac{\sum_v^N\theta\qua*{(\nu+1)D_v+M_0-i}\frac{\rho N_{vi}+1}{\rho t_u + (\nu+1) D_v + M_0}}{N^2}.
\]
If we denote by $D_u(t)$ the number of distinct items clicked by user $u$ at time $t$, the probability of $D_u$ to increase by a unit in a single time step is then obtained by summing the $R_{ui}$ over all the items $u$ has not clicked yet, so for which $N_{ui}=0$. We denote by $P_{D_u}$ the probability obtained summing these transition rates:
\[
    P_{D_u}=\sum_{i=D_u}^MR_{ui},
\]
where $M=\max_u\qua*{M_u}=\max_u\qua*{(\nu+1)D_u+M_0}$. In these terms, we thus have:
\[
    D_u(t+\delta t)=D_u(t)+P_{D_u} \ \to \ \frac{dD_u(t)}{dt}= \frac{1}{\delta t}P_{D_u}=NP_{D_u}.
\]
Let us focus on the right side of the differential equation. We can write it as
\[
    NP_{D_u}=\sum_{i=D_u}^M\frac{\sum_v^N\theta\qua*{(\nu+1)D_v+M_0-i}\frac{\rho N_{vi}+1}{\rho t_u + (\nu+1) D_v + M_0}}{N}.
\]
To simplify the expression, we make some approximation: we set $t_u\approx t$, we neglect $M_0$, and we assume $D_u$ to scale sublinearly so that it is sub-dominant compared to $t_u$. This gives 
\begin{align*}
    NP_{D_u}&=\frac{1}{N\rho t}\sum_{i=D_u}^M\sum_v^N\theta\qua*{(\nu+1)D_v-i}(\rho N_{vi}+1)\\
    &=\frac{1}{N\rho t}\gra*{\sum_{i=D_u}^{M_u}1+\sum_{v\neq u}^N\sum_{i=D_u}^{M_v}(\rho N_{vi}+1)},
\end{align*}
where we assumed there are no major fluctuations between users and so $D_u<(\nu+1)D_v$. Considering that $M_u=(\nu+1)D_u+M_0\approx(\nu+1)D_u$, this expression can be rewritten as 
\begin{align*}
    NP_{D_u}&=\frac{1}{N\rho t}\gra*{\nu D_u+\sum_{v\neq u}^N\qua*{\sum_{i=D_u}^{M_v}\rho N_{vi}+\sum_{i=D_u}^{M_v}1}}\approx\\
    &\approx\frac{1}{N\rho t}\left\{\nu D_u+\sum_{v\neq u}^N\left[\theta\qua*{D_v-D_u}\sum_{i=D_u}^{D_v}\rho\right.\right.+\\
    &\hspace{10em} +(\nu+1)D_v-D_u\Biggr]\Biggr\}.
\end{align*}
Here we made the approximation $N_{vi}=1$ for those items that have already been clicked by user $v$ but not by $u$, thus neglecting the possibility of multiple clicks. Concluding we have
\begin{align*}
    NP_{D_u}&\approx\frac{1}{N\rho t}\Bigl\{\nu D_u+\sum_{v\neq u}^N\Bigl[\theta\qua*{D_v-D_u}(D_v-D_u)\rho+\\
    &\hspace{10em} +\nu D_v-(D_u-D_v)\Bigr]\Biggr\}.
\end{align*}
and putting everything back together, we finally get
    \begin{widetext}
            \begin{equation}
                \frac{dD_u}{dt}\approx\frac{1}{N\rho t}\gra*{\nu D_u+\sum_{v\neq u}\qua*{\nu D_v-(D_u-D_v)+\rho(D_v-D_u)\theta\ton*{D_v-D_u}}},
                \label{eq:evolution_Du_general}
            \end{equation}
    \end{widetext}
The first term is the single user contribution and gives the probability to select a new item directly from the adjacent possible of the user itself. The summation instead considers the interaction with other users through the collaborative filtering mechanism and quantifies the probability of selecting an item contained in the adjacent possible of other agents. The first contribution quantifies the items in $v$'s adjacent possible. However, we have to consider that some of such items will not be novelties for $u$, and we do so through the second term. Finally, the last contribution takes into account that if $D_v>D_u$, some of the elements already clicked (and reinforced) by $v$ are still in the adjacent possible of $u$ \footnote{Note that here we are neglecting the possibility of these items to be clicked more than once.}. To simplify the problem, we assume the distribution of $D_u$ to be symmetric, and we estimate the difference $D_v-D_u$ as the standard deviation of $D_u$, that we know to satisfy Taylor's law in the UMT $\sigma(D_u)\approx \mean*{D_u}$~\cite{tria2018zipf}. In this way, defining $D\equiv\mean*{D_u}$ and taking the average value, one obtains:
\[
    \frac{dD}{dt}\approx\frac{1}{N\rho t}\qua*{\nu D+(N-1)\nu D+\frac{N-1}{2}\rho D},
\] 
where we also made the assumption $N\gg1$, and we used that the second term in the summation averages to zero. This gives:
\[
\frac{dD}{dt}\approx\frac{2\nu+\rho}{2\rho}\frac{D}{t} \to D(t)\sim t^{\frac{2\nu+\rho}{2\rho}}.
\]
The conclusion is that also in the presence of the collaborative filtering algorithm, the system shows Heaps' law with an exponent $\beta_{CF}$ satisfying the following relation:
\begin{equation}
    \beta_{CF}\approx\frac{2\nu+\rho}{2\rho}=\beta_{UMT}+\frac{1}{2}\geq\beta_{UMT},
    \label{eq:heaps_exponent_CF}
    \end{equation}
where $\beta_{UMT}$ is the exponent for the single user or, equivalently, in the absence of the recommendation algorithm. We thus see that thanks to the recommender system, agents experience novelties faster than the single user case, and we thus recovered the results of the numerical simulations. Also, note that the exponent is independent of $N$, provided that $N$ is sufficiently large, and this implies that the pace of novelties is stable even when new users enter the platform; this is a crucial requirement for online platforms where new users constantly enter or leave. In our computations, we always assumed $D(t)$ to scale sublinearly, so we have to check the value $\rho_c$ for which Heaps' exponent approaches one. It is easy to see that it holds $\rho_c=2\nu$ and that, for $\rho<\rho_c$, Eq.~\eqref{eq:heaps_exponent_CF} predicts a Heaps' exponent larger than one. However, novelties can grow most linearly in time (provided one considers the intrinsic time in which one unit of time corresponds to one event), so we expect Heaps' exponent $\beta_{CF}$ to satisfy: 
\begin{equation}
            \beta_{CF}=
            \begin{cases}
                \frac{\nu}{\rho}+\frac{1}{2}\ \text{for} \ \rho>2\nu\\
                1\ \text{for} \ \rho\leq2\nu\\
            \end{cases}
            \label{eq:heaps_exponent_CF_full}
\end{equation}
This result explains why the linear regime lasts for higher reinforcement values than the single-user scenario in the presence of recommendation algorithms. We conclude by comparing the predicted Heaps' exponent with that observed in numerical simulations. As it is possible to see in Fig.~\ref{fig:theoretical_predictions}, there are some discrepancies between theory and simulations. Still, our computations agree with the empirical Heaps' exponent, $\beta$, provided that the ratio $\nu/\rho$ is not too large. While we predict $\beta \to 0.5$ for $\rho \to \infty$, we observe the scaling exponent to converge to smaller values (but not to zero!) in this limit. Note that fluctuations among different simulations become very strong when $\rho$ gets substantially large. Very likely, in this regime, it is impossible to approximate the stochastic dynamics with deterministic differential equations because random fluctuations start to play a role. 
\section{Conclusions}
    
In the present paper, we have presented a theoretical and numerical analysis of how recommendation algorithms affect novelty discovery and the exploration process of users on online platforms. Our results show that recommender systems may increase the pace of novelties' discovery compared to the independent users' case. We tested this conclusion considering different recommendation scenarios, finding no substantial differences in the rate of novelties' discovery. On the other hand, different recommendation schemes can lead to different outcomes in terms of audience polarization. For instance, the matrix factorization algorithm induces opinion polarization compared to the user-user collaborative filtering. This evidence suggests that the latter should be preferred and that approximated techniques, despite their better computational performances in real scenarios, might have potentially dangerous unforeseen consequences on users' opinion dynamics. The framework introduced opens new possibilities in studying recommendation algorithms and their effects on individual users and society, allowing us to compare them and determine their strengths and weaknesses. Further analysis could involve investigating more advanced recommendation algorithms, such as those based on artificial neural networks now adopted by most online giants~\cite{covington2016deep, ying2018graph}. Also, the effects of social interaction among users, as on online social networks, is an interesting research avenue. For instance, one could consider both the impact of a social filtering algorithm~\cite{tang2013social} or the presence of an underlying interaction network~\cite{iacopini2020interacting}, which has been proven to enhance novelties. We believe our work will pave the way for the study and better comprehension of recommendation algorithms, opening the development of more sustainable recommendation techniques beyond the standard accuracy-based metrics.

\end{document}